\documentclass[sigconf]{acmart}
\usepackage{multirow}
\usepackage{hyperref}
\usepackage{xcolor}
\usepackage{pifont}
\usepackage{colortbl}
\usepackage{adjustbox}
\usepackage{array, makecell}
\usepackage{tikz}
\usepackage{tabularx}
\usepackage{hyperref}
    \newcolumntype{L}{>{\raggedright\arraybackslash}X}
\AtBeginDocument{%
  \providecommand\BibTeX{{%
    \normalfont B\kern-0.5em{\scshape i\kern-0.25em b}\kern-0.8em\TeX}}}

\setcopyright{acmcopyright}
\copyrightyear{2018}
\acmYear{2018}
\acmDOI{XXXXXXX.XXXXXXX}

\acmConference[Conference acronym 'XX]{Make sure to enter the correct
  conference title from your rights confirmation emai}{June 03--05,
  2018}{Woodstock, NY}
\acmPrice{15.00}
\acmISBN{978-1-4503-XXXX-X/18/06}

\begin{document}

\title{Workshop on Document Intelligence Understanding\\
\url{https://doc-iu.github.io/}}




\begin{abstract}
  Document understanding and information extraction include different tasks to understand a document and extract valuable information automatically. Recently, there has been a rising demand for developing document understanding among different domains, including business, law, and medicine, to boost the efficiency of work that is associated with a large number of documents. 
  
  This workshop aims to bring together researchers and industry developers in the field of document intelligence and understanding diverse document types to boost automatic document processing and understanding techniques. We also release a data challenge on the recently introduced document-level VQA dataset, PDFVQA\footnote{\url{https://www.kaggle.com/t/ce5b8d5610c24d719d9a76020700f8bf}}. The PDFVQA challenge examines the model’s structural and contextual understandings on the natural full document level of multiple consecutive document pages by including questions with a sequence of answers extracted from multi-pages of the full document. This task helps to boost the document understanding step from the single-page level to the full document level understanding. 
\end{abstract}



\keywords{Document Understanding, Information Extraction, Layout Analyzing, Visual Question Answering}


\maketitle

\section{Workshop Organizers}
\textbf{Dr Caren Han}
\\The University of Sydney and The University of Western Australia, Australia, \href{mailto:}{caren.han@sydney.edu.au} and \href{mailto:}{caren.han@uwa.edu.au}

\noindent \textbf{Mr. Yihao Ding} \\The University of Sydney, Australia, \href{mailto:}{yihao.ding@sydney.edu.au}

\noindent \textbf{Ms. Siwen Luo}
\\The University of Sydney and The University of Western Australia, Australia, \href{mailto:}{siwen.luo@sydney.edu.au} and \href{mailto:}{siwen.luo@uwa.edu.au}

\noindent \textbf{Dr. Josiah Poon} \\The University of Sydney, Australia, \href{mailto:}{josiah.poon@sydney.edu.au}

\begin{figure}[t]
 \centering
 \includegraphics[width=0.98\linewidth]{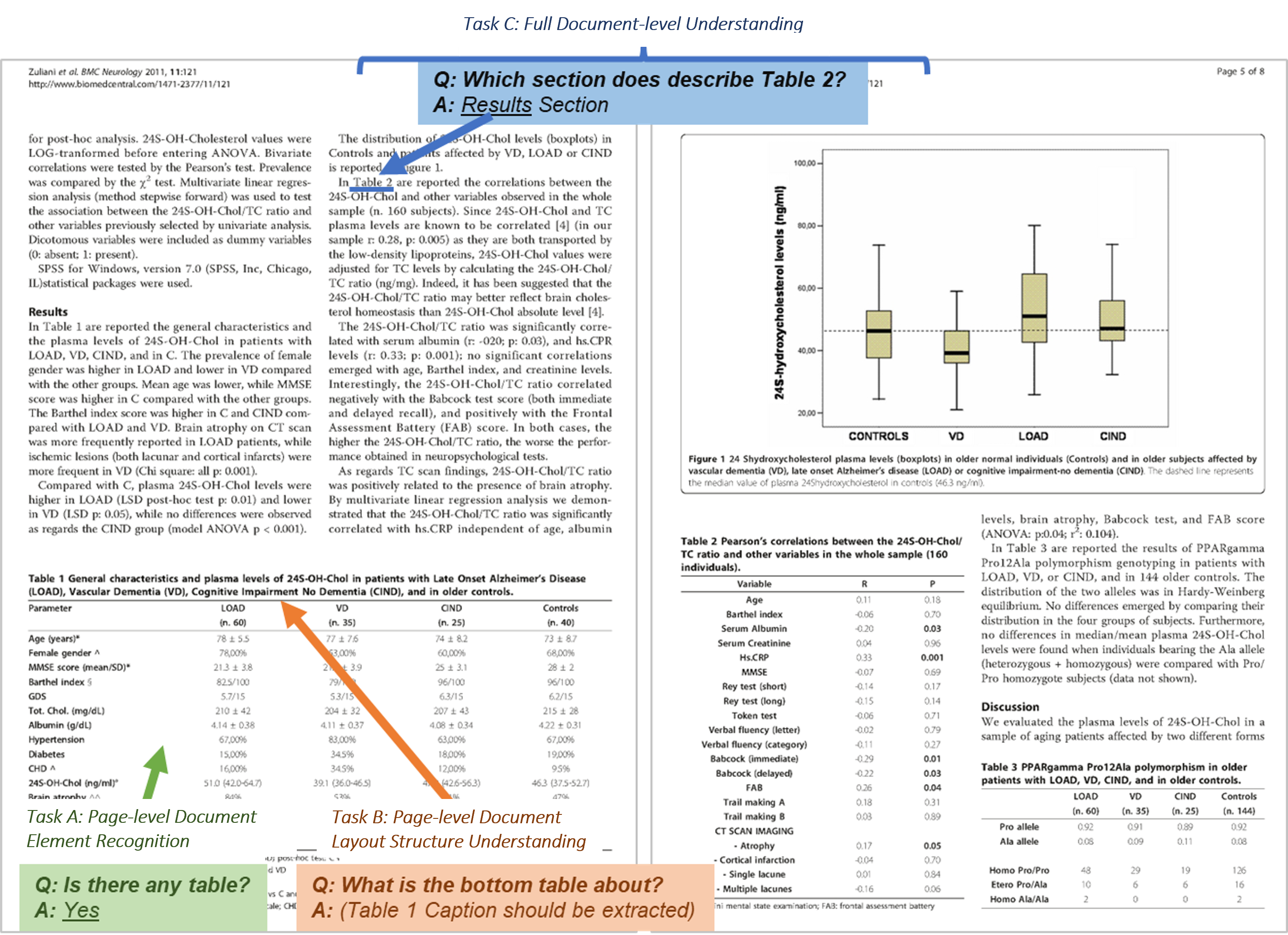}
 \caption{PDF-VQA Data Challenge Sample Questions and Document Pages for Task A, B, and C. \cite{pdfvqa}}
 \label{fig:system}
\end{figure}

\section{Workshop Contact Person}
\textbf{Dr. Caren Han}
\\\textbf{Address:} The University of Western Australia (M002), 35 Stirling Highway, 6009 Perth, Australia
(+61 423915170)
\\\textbf{Email:} \href{mailto:}{caren.han@uwa.edu.au}
\\\textbf{Website:} \url{https://drcarenhan.github.io/}

\begin{table*}[t]
    \centering
    \begin{tabularx}{\linewidth}{|L|l|l|l|}
    \hline
        \textbf{Workshop Name} & \textbf{Conference} & \textbf{Year} & \textbf{Website}\\
    \hline
    \multicolumn{4}{|c|}{\textbf{Document Intelligence or Processing Workshops}}\\
    \hline
        DocVQA & ICDAR & 2021 & \url{https://www.docvqa.org/workshops}\\
        Scholarly Document Processing & EMNLP, NAACL, COLING & 2020, 2021, 2022 & \url{https://sdproc.org/2022/index.html}\\
        Document Intelligence & KDD & 2022 & \url{https://document-intelligence.github.io/DI-2022/}\\
    \hline
    \multicolumn{4}{|c|}{\textbf{Information Retrieval Workshops (including document understanding aspect)}}\\
    \hline
    Proactive And Agent-Supported Information Retrieval & CIKM & 2022 & \url{https://pasircikm2022.github.io/PASIRCIKM/}\\
    Federated Learning for Information Retrieval & SIGIR & 2023 & \url{https://sites.google.com/view/flirt-sigir23}\\
    \hline
    \end{tabularx}
    \caption{Related Workshop List in Previous Conferences.}
    \label{tab:related_workshop}
\end{table*}

\section{Workshop Theme and Topics}
This workshop aims to explore and advance the current state of research and technical articles, including but not limited to the topics. Note that the models for the data challenges should apply the following technology:

\noindent\textbf{Document Processing and Structure Understanding:} Benchmark datasets, models, and off-the-shelf applications for processing unstructured or semi-structured document images into machine-readable formats, including but not limited to:
\begin {itemize}
    \item Document Layout Analysis
    \item Document Parsing
    \item Document Image Processing
    \item Document Table Detection
    \item Table Structure Recognition
    \item Reading Order Prediction
\end{itemize}
\noindent\textbf{Document Content Understanding:} Benchmark datasets, models and off-the-shelf applications focusing on understanding the content of target documents to extract critical information and conducting downstream analysis:
\begin {itemize}
    \item Document Visual Question Answering
    \item Document Information Retrieval
    \item Document Key Information Extraction 
    \item Document Classification and Categorization
    \item Historial Document Content Understanding
    \item Table-based Question Answering
\end{itemize}
\noindent\textbf{Advanced Deep Learning Approaches for Doc-IU:} Recently proposed deep learning techniques boosting the development of document intelligence including but not limited to:
\begin {itemize}
    \item Document Feature Representation
    \item Multi-modal Fusion and Adaptive Learning
    \item Pre-training Mechanisms for Document Understanding
    \item Few-shot Learning and Zero-shot Learning
\end{itemize}

\section{Workshop Objectives, Goals \& Expected Outcome}
This workshop aims to invite and share recent research works and technical reports of various document understanding tasks, including document layout analysis\cite{publaynet,docbank,docgcn,vdoc}, document visual question answering\cite{pdfvqa,docvqa,visualmrc}, document key-value extraction\cite{cord,sroie}. It provides a discussion panel for researchers in this field to share and discuss the research work trend and techniques for better document understanding. 

Moreover, this workshop proposes a document VQA data challenge with the PDFVQA dataset (Challenge website: \url{https://www.kaggle.com/competitions/pdfvqa}). This challenge aims to develop models to answer the questions with the given document images. Unlike the other document VQA datasets that focus on the contextual understanding of texts, PDFVQA questions target the structural relationship understandings among the document layout components. Moreover, instead of processing on independent document images, PDFVQA datasets contain the whole document of multiple pages, and there could be multiple answers that span over multiple pages for one question. PDFVQA motivates the document understanding and processing not limited to the single document page but expands to the full document level to understand the logic and connections of document contents and structures over consecutive pages. We expect through the proposed challenge, researchers in document understanding would pay more attention to and develop new techniques for the document understanding task on the full document level. 

As shown in Figure\ref{fig:system}, answering a question requires reviewing the full document contents and identifying the contents hierarchically related to the queried item in the question. For example, the question ``Which section does describe Table 2?" in Figure~\ref{fig:system} requires the identification of all the sections of the full document that have described the queried table. The answers to such questions are the texts of the corresponding section titles extracted as the high-level summarization of the identified sections. Identifying the items at the higher-level hierarchy of the queried item is defined as the parent relation understanding the question in PDF-VQA. Oppositely, Task C also contains the questions of identifying the items at the lower-level hierarchy of the queried item, and such questions are defined as the child relation understanding. For example, a question, ``What does the `Methods' section about?" requires extracting all the subsection titles as the answer.

\section{Workshop Length}
The workshop will be hosted in half-day (4 hours\footnote{This can be updated based on the decision made by the CIKM workshop chairs}), containing three main sessions. Each session is scheduled to be roughly 1 hour (60 mins), with two 15-min breaks in between. The PDF dataset challenge session would include the Winners' presentation and the Award Ceremony. The Award Ceremony will be sponsored by \href{https://research.google/}{Google Research} and \href{https://www.fortifyedge.com/}{FortifyEdge}. The detailed length and schedule are in Table \ref{tab:schedule}.
\begin{table*}[t]
    \centering
    \begin{tabularx}{\linewidth}{|l|l|L|}
    \hline
       \textbf{Length} & \textbf{Invite Speaker} & \textbf{Topic} \\
    \hline
      10 mins & Dr. Caren Han & Workshop Opening Speech \\ 
    \hline
         \multicolumn{3}{|c|}{\textbf{Document Understanding Datasets Session}}\\
    \hline
      60 mins & \multicolumn{2}{|c|}{\makecell[c]
{\\ Document Understanding Benchmark Dataset papers published in \\ CIKM, SIGIR, KDD, ICDM, ACL, EMNLP and NAACL conferences 2021-2023 \\ \\ }}\\

    \hline
    15 mins   & \multicolumn{2}{|c|}{\textbf{break}}\\
    \hline
    \multicolumn{3}{|c|}{\textbf{Document Understanding Model Session}}\\
    \hline
    60 mins & \multicolumn{2}{|c|}{\makecell[c]
{\\ Document Understanding Model papers published in top-tier \\ CIKM, SIGIR, KDD, ICDM, ACL, EMNLP and NAACL conferences 2021-2023 \\ \\ }}\\

    \hline
    15 mins & \multicolumn{2}{|c|}{\textbf{break}}\\
    \hline
        \multicolumn{3}{|c|}{\textbf{PDFVQA Challenge Session}}\\
    \hline  
     10 mins  & Mr. Yihao Ding & Leaderboard Opening \\
    \hline
     20 mins  & Ms. Siwen Luo & PDF-VQA: A New Dataset for Real-World VQA on PDF Documents \\ 
    \hline
     30 mins  & Winner & Challenge Winner Presentation on Methodology and Results \\
    \hline
     10 mins  & Dr. Josiah Poon & Awards Ceremony \\
    \hline
     10 mins  & Dr. Caren Han & Conclusion Remarks \\
    \hline
    \end{tabularx}
    \caption{Workshop Program Schedule}
    \label{tab:schedule}
\end{table*}

\section{Target Audience}
The workshop would be beneficial to researchers, industrial developers, and practitioners who are interested in broad information retrieval, knowledge management, natural language processing, machine learning, and data mining, specifically document understanding, document intelligence, and document content understanding. While the audience with a good background in the above areas would benefit most from this workshop,  we believe that the presented research keynote speech, and model architectures and technical details to address the dataset challenge would give the general audience and newcomers a complete picture of the current work and inspire them to learn more in this field. 
The workshop is designed as self-contained, so no specific background knowledge is assumed of the audience. However, it would be advantageous for the audience to know about basic multi-modal deep learning technologies and new applications of large pre-trained models on documents (a new type of image source). We will provide the audience with the reading list and dataset samples on our workshop website. Note that the workshop website is provided in \url{https://doc-iu.github.io/}.

\section{Workshop Relevance}
The workshop is highly relevant to the CIKM community on research on information and knowledge management. It also directly aligned with the aim of the CIKM workshops, bridging the academic-commercial gap in the database, information retrieval, machine learning, and knowledge management communities. As informed in the CIKM workshop proposal website, the proposed workshop, DocIU, would enable participants to propose novel research and present practical applications on document intelligence and understanding by addressing the research problem of the dataset challenge, PDF-VQA. It would require many aspects of the data lifecycle, including acquisition, pre-processing, modelling, integration/aggregation, and analysis. 
The workshop finally related to the aim of the CIKM workshop, providing interdisciplinary workshops bridging across different communities. The aim of the proposed workshop and workshop organizers have rich interdisciplinary aspects in major Natural Language Processing, Information Retrieval, Computer Vision, and Knowledge Management.

\section{Related Workshops}
The workshop is considered a cutting-edge workshop that covers the recent trends in emerging areas of information retrieval, knowledge management, and natural language processing. As shown in Table \ref{tab:related_workshop}, previous workshops in top-tier Natural Language Processing conferences, including EMNLP, NAACL, COLING, and KDD, covered similar but not the same topics in document intelligence and processing. The DocVQA workshop was only limited to the Document Visual Question Answering, while the scholarly Document Processing workshop has a broader document processing topic range, not focusing on the multimodal document image and text processing as our proposed workshop. Their topics focus more on text processing, including generation and summarization. Document Intelligence workshop focuses more on document understanding tasks. But their topics are limited to business document understanding while our workshop encourages broader exploration of diverse document types.

The CIKM or SIGIR workshops have covered information retrieval or federated learning using document types but have not touched on the exact document understanding or intelligence in the past three years. The workshop has not been released elsewhere.



\section{Workshop Program Format}
Workshop chairs Dr. Caren Han and Ms. Siwen Luo and Leaderboard Chair Mr. Yihao Ding will organize the workshop in person. The workshop contains three main sessions. The first session includes the works to introduce different document understanding tasks and the benchmark dataset works. The second session includes the works of key document understanding models. The last session will be the PDFVQA challenge session. The challenge winner will present the methodology and results, followed by a QA session. The detailed workshop program is listed in Table \ref{tab:schedule}.

\section{Workshop Schedule \& Important Dates}
\begin{itemize}
    \item Leaderboard Challenge Due: 15 September 2023
    \item Workshop Paper Submission Due: 15 September 2023
    \item Announcement of Winner: 30 September 2023
    \item Paper Acceptance notification: 30 September 2023
    \item Workshop Date: 22 October 2023
\end{itemize}

\section{Program Committee}
\begin{itemize}
    \item Workshop Chair: Dr. Caren Han, The University of Sydney \& The University of Western Australia, Australia
    \item Workshop Chair: Mr. Yihao Ding, The University of Sydney, Australia
    \item Workshop Chair: Ms. Siwen Luo, The University of Sydney \& The University of Western Australia, Australia
    \item Workshop Chair: Dr. Josiah Poon, The University of Western Australia, Australia
    \item Advisory Committee: Mr. Zhe Huang, ANT Group \& Alibaba Group, China 
    \item Advisory Committee: Dr. HeeGuen Yoon, National Information Society Agency, Korea
    \item Advisory Committee: Dr. Paul Duuring, Department of Mines and Petroleum, Australia
    \item Advisory Committee: Prof. Eun-Jung Holden, UWA Institute of Data, Australia
\end{itemize}

\section{Participation \& Selection Process}
Participants are selected with their works in the document understanding domain published at top-tier Information Retrieval and Natural Language Processing conferences, including CIKM, SIGIR, KDD, ICDM, ACL, EMNLP and NAACL. 

For the PDFVQA challenge, the winner will be chosen via the top leaderboard score. In addition, challenge participants must submit all technical reports, including the code and a 2-page report illustrating the methodology and testing results. The winner selection criteria also include the validity and reproductivity of the submitted code and the quality of the report.\footnote{Due to the Award Prize transfer, note that attendees from Countries in Sanctions list will not be considered}.

\section{Organizers' Background}
The workshop organizers' are all experts in Natural Language Processing and Information Retrieval, and the detailed background information can be found as follows:

\noindent\textbf{Mr. Yihao Ding (In person)} is a PhD candidate at the School of Computer Science, University of Sydney, and visiting scholar at the School of Computer Science, University of Western Australia. He is a Research Assistant at the School of Earth Science, University of Western Australia. He received his Bachelor's Degree and Master's Degree in 2015 and 2018 in Geospatial Engineering and his second Master's Degree in Information Technology in 2020. His research interests include deep learning-based document analysis, information retrieval, and graph neural networks. He has published several top-tier conferences and journal papers in CVPR, SIGIR, COLING, ECML-PKDD, and Frontiers. 

\noindent\textbf{Ms. Siwen Luo (In person)} is a final year PhD student at the School of Computer Science, The University of Sydney, and a visiting scholar at the School of Physics, Mathematics and Computers, University of Western Australia. Her research focuses on the cross-area of computer vision and natural language processing, aiming for the exploration and development of interpretable models for multimodalities. Her research spans a range of multimodal tasks, including Visual Question Answering, Text to Image generation, and Document Layout Analysis. She has published several papers at top-tier NLP conferences and received the Best Paper Award from ICONIP 2020 and the best area paper from COLING 2020. 
    
\noindent\textbf{Dr. Soyeon Caren Han (In person)} is a co-leader of AD-NLP (Australia Deep Learning NLP Group) and a Senior lecturer (Associate Professor in U.S. System) at the University of Western Australia and an honorary senior lecturer (honorary Associate Professor in U.S. System) at the University of Sydney and the University of Edinburgh. After her Ph.D.(in 2017), she worked for six years at the University of Sydney. Her research interests include Natural Language Processing with Deep Learning. She is broadly interested in several research topics, including visual-linguistic multimodal learning, abusive language detection, document analysis, and recommender system. More information can be found at \url{https://drcarenhan.github.io/}.
    
\noindent\textbf{Dr. Josiah Poon} is a co-leader of AD-NLP (Australia Deep Learning NLP Group) and a Senior Lecturer at the School of Computer Science, University of Sydney. He’s been using traditional machine learning techniques paying particular attention to learning from imbalanced datasets, short string text classification, and data complexity analysis. He has coordinated a multidisciplinary team consisting of computer scientists, pharmacists, western medicine \& traditional Chinese medicine researchers and practitioners since 2007. He co-leads a joint big-data laboratory for integrative medicine (Acclaim) established between the University of Sydney and the Chinese University of Hong Kong to study medical/health problems using computational tools.

\subsection{Advisory Committees}
We also invite four advisory committees from diverse domain backgrounds and different countries: 

\noindent\textbf{Dr. HeeGuen Yoon}, National Information Society Agency, \textbf{Korea}

\noindent\textbf{Mr. Zhe Huang}, ANT Group, Alibaba Group, \textbf{China}

\noindent\textbf{Dr. Paul Duuring}, Department of Mines and Petroleum, \textbf{Australia}

\noindent\textbf{Prof. Eun-Jung Holden}, UWA Institute of Data, \textbf{Australia}

\bibliographystyle{ACM-Reference-Format}
\bibliography{sample-base}

\end{document}